\documentclass[showpacs,oneside,twocolumn,prl,amsmath,amssymb,fleqn]{revtex4-1}
\usepackage{cases}
\usepackage{amsmath}
\usepackage{amssymb}
\usepackage{amsfonts}
\usepackage{amssymb}
\usepackage{dcolumn}
\usepackage{bm}
\usepackage{bbm}
\usepackage{graphicx}
\usepackage{xcolor}
\usepackage{array}
\usepackage{subfigure}
\usepackage[none]{hyphenat}
\usepackage{ulem}

\begin{document}
\title{A Maxwell demon model connecting information and thermodynamics}
\author{Peiyan Peng}
\author{Changkui Duan}

\altaffiliation{yanyuan@mail.ustc.edu.cn}
\affiliation{Hefei National Laboratory for Physics Sciences at
Microscale and Department of Modern Physics, University of Science
and Technology of China, Hefei, 230026, China}

\begin{abstract}
In the past decade several theoretical Maxwell's demon models have been proposed exhibiting effects such as refrigerating, doing work at the cost of information, and some experiments have been done to realise these effects. Here we propose a model with a two level demon, information represented by a sequence of bits, and two heat reservoirs. Which reservoir the demon interact with depends on the bit. If information is pure, one reservoir will be refrigerated, on the other hand, information can be erased if temperature difference is large. Genuine examples of such a system are discussed.
\end{abstract}

\maketitle

  Maxwell's demon as an example that violates the second law of thermodynamics was not solved for a long time, since people used to focus on the thermodynamic system only and neglect the information cost of the controlling process of the demon. The study into the energy cost of information erasing process lead to the Landauer's principle \cite{text13}: To erase one bit of information, the energy cost is at least ln2$k_B T$. Here $T$ stands for the temperature of the information storage environment. This principle bridges information theory and thermodynamics, and helped to solve the paradox eventually \cite{text10,text11,text9}. Although the thermodynamic system alone may decrease in entropy with the help of Maxwell's demon, but when taken into account the storage of information, i.e, the increase of information entropy, the overall system's entropy would not increase. On the other hand, the free energy increasement gained by the measurement can not compensate for the information erasing cost except for reversible process. So the second law of thermodynamics is recovered. A more recent result is the generalised Jarzynski equality \cite{text15,text17,text18,text16}which contains a mutual information part quantifying the information gained by Maxwell demon during thermodynamic process. When information are represented by bits, considering the concavity of exponential function, the result of Landauer's principle is recovered \cite{text9,text19}.

  These perspective make it possible to design heat engines that work just like the Maxwell's demon \cite{text1,text2,text5,text6,text7,text8,text20,text23,text26,text28}, which stores information and can refrigerate thermal systems or do work. Notice that information entropy increases in the information storing process, while erasing information reduces the information entropy. On quantum mechanic level, NMR system Maxwell's demon was proposed to refrigerate the temperature of nuclear spins \cite{text3}. Upon former models \cite{text1,text7,text23}, we propose a simple thermodynamic system with a two-level quantum system as Maxwell's demon. We analysed the thermodynamics of the process, calculated the relative efficiency to confirm the second law of thermodynamics. The possible realisations in real systems are considered as well.

  In our proposal, the Maxwell's demon is a simple two level system, the demon's energy level can be coupled to information, here the information means a sequence of bits. One single bit is coupled to the demon at a time, which may be 0 or 1. This model also includes two heat reservoir with different temperatures $T_1$ and $T_2$. When the demon is in the up level and the bit is 1, the demon may release an amount of energy $\Delta E$ to $T_1$, and the information turns to 0. The rate of transition is determined by the $T_1$. When the demon is in the up level and the bit is 0, $\Delta E$ may be released to the $T_2$, after transition the bit changes to 1. Both transitions are reversible and transition possibilities satisfy detailed balance.

\begin{figure}
\includegraphics[width=3.0in,height=1.50in]{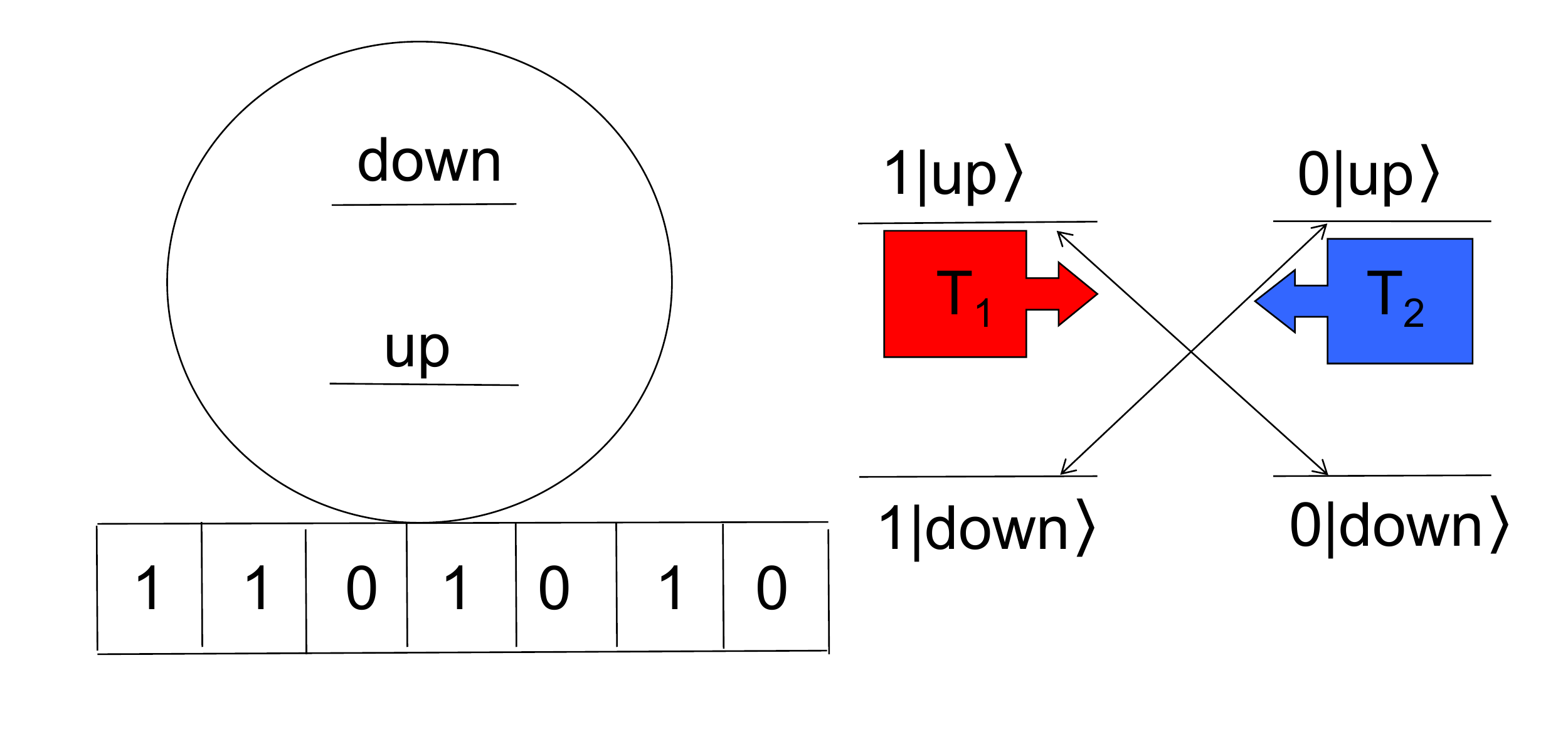}

\caption{ A two level Maxwell demon and a sequence of classical bits. Which reservoir the demon interact with is determined by the bit.}
\label{fig1}
\end{figure}

  We denote the proportion of 0 for the incoming information as $p$, the probability of the demon being in the up state is $q$, and initially the demon and the bit has no coupling. The united distribution $ (1u,1d,0u,0d)=\{(1-p)q,(1-p)(1-q),pq,p(1-q)\}$. The transitions in the two thermal reservoirs are independent of each other, so the two allowed transitions satisfy their sperate probability conservation:
  \begin{eqnarray}
  P_{0u}(t)+P_{1d}(t)&=&P_{0u}(0)+P_{1d}(0),\nonumber\\P_{1u}(t)+P_{0d}(t)&=&P_{1u}(0)+P_{0d}(0).
  \end{eqnarray}
Here $P_{0u}(t)$, $P_{1u}(t)$, $P_{0d}(t)$, and $P_{1d}(t)$ stand for the population of states at time. After coupling the four states are not in thermodynamic equilibrium initially, they can exchange energy with the two heat reservoir to reach a stable state. We assume that the interacting process is long enough so that the four states can eventually reach the statistical equilibrium state. According to thermodynamics laws, the final states satisfy Boltzman distribution. That is :
  \begin{eqnarray}
\frac{P_{0u}}{P_{1d}}&=&\exp (-\beta _2\Delta E),\nonumber \\\frac{P_{1u}}{P_{0d}}&=&\exp (-\beta _1\Delta E)
  \end{eqnarray}
  where $\beta _1=\frac{1}{k_B T_1}$ and $\beta _2=\frac{1}{k_B T_2}$. Given the initial distribution, the four conditions are enough to calculate the final probabilities distribution.
  \begin{eqnarray}
  \left(\begin{array}{c}1u\\1d\\0u\\0d\end{array}\right)&=&\left(\begin{array}{c}\frac{(p+q-2pq)\exp (-\beta _1\Delta E)}{1+\exp (-\beta _1\Delta E)}\\\frac{(1-p-q+2pq)}{1+\exp (-\beta _2\Delta E)}\\ \frac{(1-p-q+2pq)\exp (-\beta _2\Delta E)}{1+\exp (-\beta _2\Delta E)}\\\frac{(p+q-2pq)}{1+\exp (-\beta _1\Delta E)}\end{array}\right).
  \end{eqnarray}Afterwards the Maxwell's demon decouples with the information, and the outgoing information is different from the incoming information because of transitions. At the same time the probability of the Maxwell's demon's state distribution changes. We can see that the demon assisted process has energy transportation and information change at the same time.

 The equilibrium state of the demon in one cycle depends on whether the incoming bit is 1 or 0. Since we have assumed the information to have a distribution, the statistical state of the demon for many cycles will be stable. This state is the periodic stable state of the demon \cite{text1,text7,text23,text12}, and can be calculated by making $q$=$P_{0u}+P_{1u}$ as given by (3). It has no reliance on the initial state and depends only on $p$ and $T_1$, $T_2$. So the demon work just like an ``information heat engine". The population of 0 and 1 after coupling can be derived from the demon state, and the amount of energy transfer can also be calculated.

 We calculate the information and energy change now. Specifically we set $T_1\geq T_2$. The amount of information is defined by entropy $S=-p\ln p-(1-p)\ln (1-p)$. The energy change of the two heat source can be calculated from the four states distribution. The amount of energy transferred from $T_2$ to $T_1$ through the demon is $\Delta Q=(P_{1u}(\infty)-P_{1u}(0))\Delta E$ on average. The entropy change of the two heat reservoirs is then $\Delta S_T=\Delta Q(\beta _2-\beta _1)$. And the information entropy change is $\Delta S_{B}=S(\infty)-S(0)$. They are both functions of $p$, $T_1$ and $T_2$. The sum of the two entropy changes is then the entropy changes of the whole system, and it is non-negative as can be seen in Fig 2.

 \begin{figure}
\centering
\includegraphics[width=2.50in,height=1.80in]{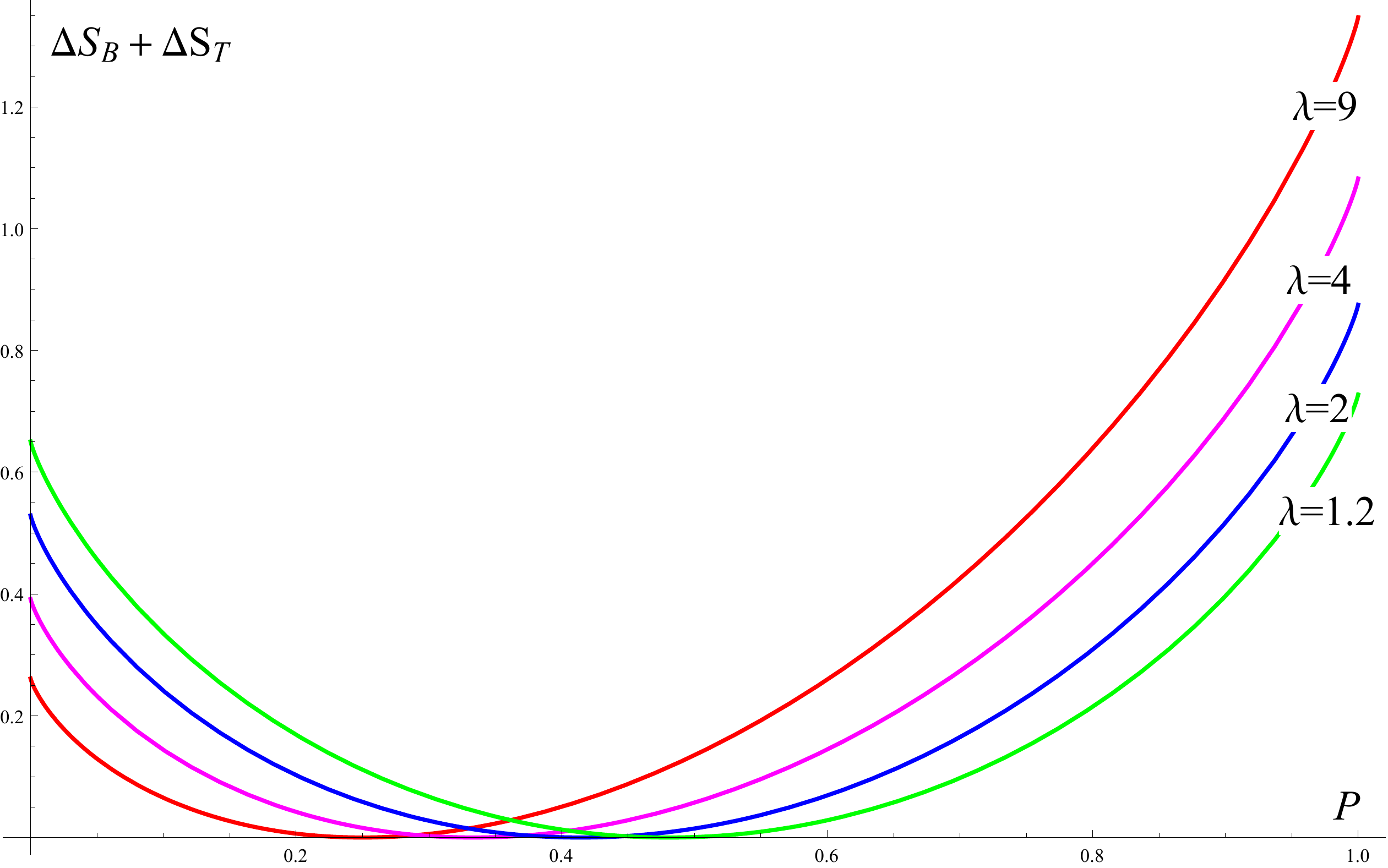}
\includegraphics[width=2.50in,height=1.80in]{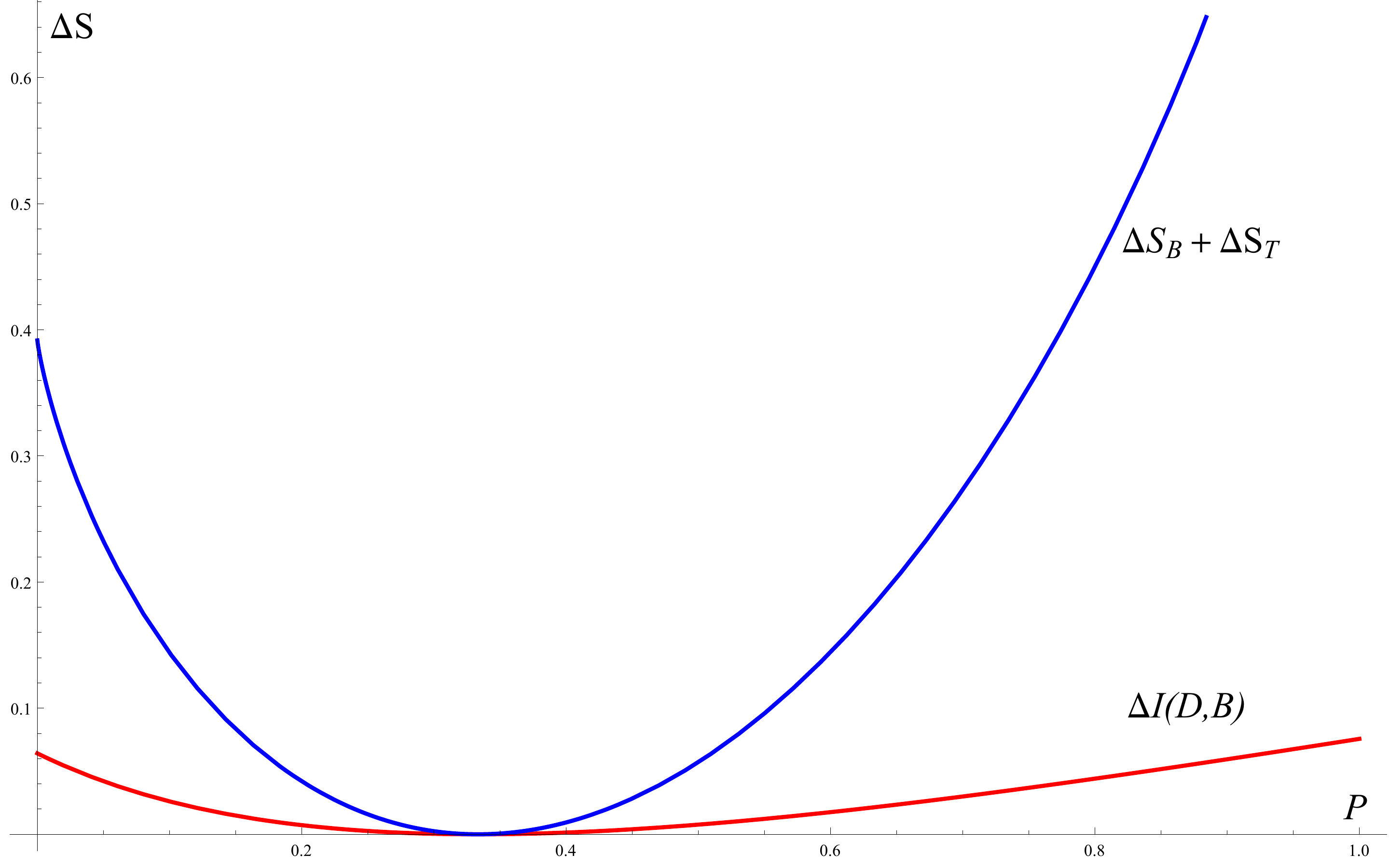}

\caption{The above picture shows the sum of the information and the two thermal reservoir's entropy change as a function of different temperatures $T_h$ and $T_c$ against $p$. We define $\lambda=\exp(T_h-T_c)\Delta E k_B$. The below picture shows the sum as well as the mutual information change of demon and bits. It is clearly seen from above curves that the entropy of the whole system never decrease during the process, so the generalised second law of thermodynamics is not violated. }
\label{fig2}
\end{figure}

 If the two temperatures are the same, we see from the two transitions that a net decrease in 0's population after interaction indicates energy transportation from $T_2$ to $T_1$. It is always possible to cool one source while heat another so long as 0 and 1 are not equally possible. Fluctuation exists for single cycle, but the overall result will not change. This is the case Maxwell initially considered. It is noticed that this is at the cost of information entropy increase.

For the case of $T_1>T_2$, the refrigerating effect depends on the information entropy. Using the critical conditions $\Delta Q=0$ and $\Delta S_{I}=0$ we get two values:
\begin{eqnarray}
p_1&=&\frac{1}{1+\exp[(\beta _2-\beta _1)\Delta E /2]},\nonumber  \\p_2&=&\frac{1}{2}(1+\tan\frac{\theta}{2}).
\end{eqnarray}
Here $\tan\theta=\frac{\exp(-\beta _1\Delta E)-\exp(-\beta _2\Delta E)}{1+\exp(-\beta _1\Delta E-\beta _2\Delta E)}$. It is clear that $p_2>1/2>p_1$. For an autonomous systems, the information erasing and the energy transportation are competitive process \cite{text1}. We plotted the change of information entropy and thermal reservoir entropy against $p$ for fixed temperature in Fig 3. It clearly shows three different regions divided by $p_1$ and $p_2$. We define $\Delta Q<0$ as refrigerating and $\Delta S_B<0$ as erasing. In the following we discuss the effects of the three regions: (a) $p<p_1$, in this regime $\Delta Q<0$ and $\Delta S_{B}>0$, so Maxwell demon absorbs energy from the lower temperature heat source, at the same time the energy is released to the higher temperature heat source. During this process the information entropy increases, while the thermal system entropy decreases. The demon works as a information engine, which can refrigerate the cold reservoir and heat the hot reservoir. A special case is when $p$ is 0, the incoming bits are all 1, the demon has only $1u$ and $1d$ initially, which means that the demon act as a infinitely high temperature source as well as a absolute zero temperature source at the same time. So according to transition rules, it is easy to see that energy has to flow from cold reservoir to hot reservoir. (b) If $p_2>p>p_1$, $\Delta Q>0$ and $\Delta S_{B}<0$, Maxwell demon gets energy from the higher temperature heat source and release it to the lower temperature heat source, decreasing the information entropy, which can be seen as the heat source doing work to erase the information. And the effects of erasing depends on the temperature difference. (c) When $p>p_2$, $\Delta Q>0$ as well as $\Delta S_{B}>0$, Maxwell demon gets energy from the higher temperature heat source and give it to the lower temperature heat source, at the same time increasing the information entropy. Take $p=1$ as a special case, the bits are all zero now, the demon's two levels now have the same temperatures as above, but transition rules lead to opposite results with respect to previous case. This is an ineffective zone. The effects are clearly dependent on temperature difference. The demon and information has no coupling in the beginning, but after one single period, the demon gains information from the bits, as indicated by the non-negativity of mutual information. As we know, the stable state of the demon is statistical, so mutual information gain for different periods has fluctuations, but on average, the generalised second law of thermodynamics is not violated. The mutual information can exceed classical counterpart and plays more important role for qubits \cite{text27}.

\begin{figure}

  \includegraphics[width=2.50in,height=1.80in]{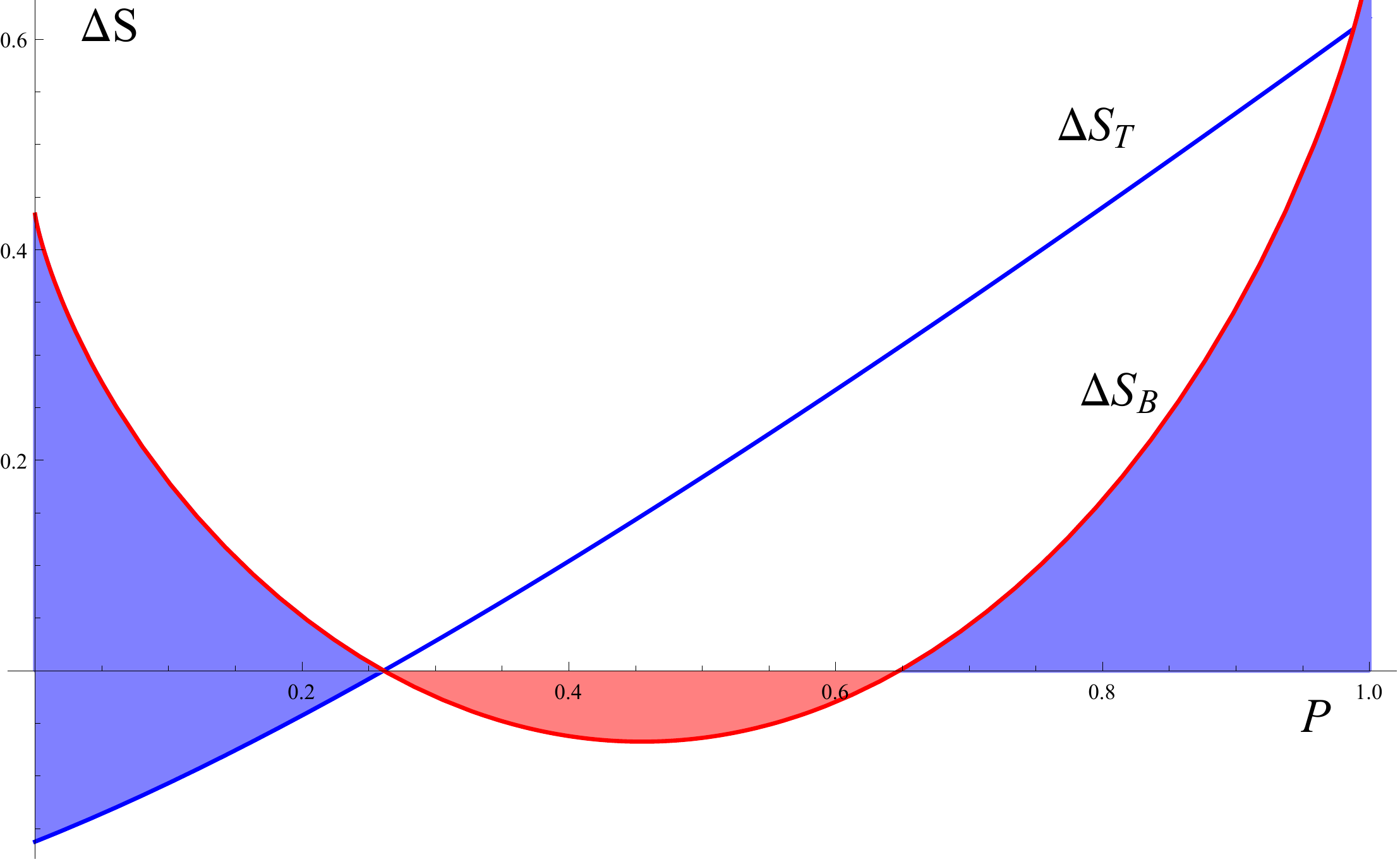}

  \caption{The entropy change effects of information and thermal reservoir as functions of $p$, here the red curve is the entropy change of the two thermal reservoir and the blue curve is the entropy change of the information. Three kinds of relations can be seen from above. The left blue area denotes refrigerating zone cause thermal entropy decreases, the red area is the erasing zone because the information entropy decreases, the right blue area is the ineffective zone. We have choosen $\exp(-\beta _h\Delta E)$ to be 0.8 and $\exp(-\beta _c\Delta E)$ to be 0.1.}
  \label{fig3}
  \end{figure}

 We see that the  overall entropy of the two heat sources and the information must not decrease during the refrigerating or erasing process. Thus we can define a relative  refrigerating efficiency as:
\begin{eqnarray}
\eta _c&=&\frac{-\Delta S_{T}}{\Delta S_{B}}.
\end{eqnarray}
 and a relative erasing efficiency when the system is in the erasing zone as:
 \begin{eqnarray}
 \eta _e&=&\frac{-\Delta S_{B}}{\Delta S_{T}}.
 \end{eqnarray}
  We plotted the relative efficiency change with respect to the initial information and the temperature difference. When temperature difference is fixed, the relative rate decreases as $p$ deviate from $p_1$. On the other hand, the erasing rate is also the highest at $p_1$, and at $p_2$ erasing drop to 0. Recall that Landauer's principle has a lower limit on the cost of entropy for erasing information, this limit is achieved at $p=p_1$. The critical probability has the highest efficiency in Fig 4 because it is the dividing point of the refrigerating zone and the  erasing zone, so no energy has transferred and the information do not change during the process, the whole system keeps still, thus we can see that this is the ideal condition. When there has nonzero energy current between the two reservoirs, the efficiency cannot be optimal, the more energy that has been transferred, the lower the efficiency is. We conclude that the ideal efficiency is reached only when the process is reversible.

\begin{figure}

  \includegraphics[width=2.40in,height=1.80in]{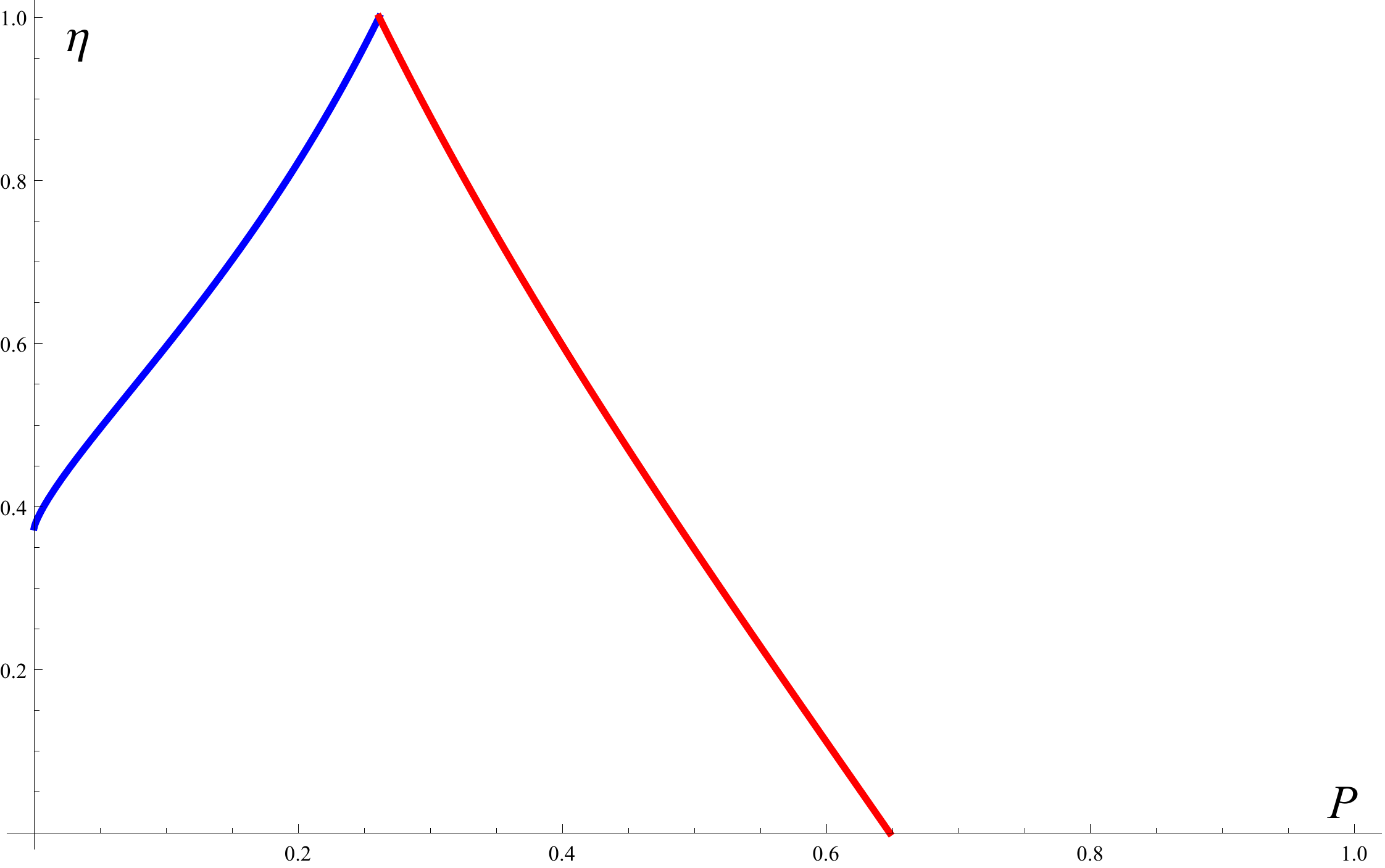}

  \caption{The relative efficiency of refrigerating and erasing against $p$ for definite temperature $T_h$, $T_c$, the red curve is the erasing efficiency, and the blue curve is the refrigerating efficiency, the point when they are both highest is the critical $p$ of refrigerating and erasing zone. We have choosing $\exp(-\beta _h\Delta E)$ to be 0.8 and $\exp(-\beta _c\Delta E)$ to be 0.1 as well.}
  \label{fig4}
\end{figure}

Here we discuss possible realisation in real systems. One example is by making use of two levels of rare-earth ions (in crystal) with Kramers degeneracy but lifted by applied weak magnetic field. If magnetic field are applied along some specific orientation, the states become distinct and have selection rules depending on the polarization of lights. One state can only absorb left-handed circularly polarized radiation and another state can only absorb right-handed circularly polarized radiation. The distribution of the pseudo-spin states stores information. The two radiation fields the rare ion interacts with can be adjusted artificially at equilibrium and can be regarded as the heat reservoirs. A schematic view is given in Fig 5. Some other systems such as the optical cavity trapped ions or 2D materials \cite{text24,text25} can also simulate our models. To find a Maxwell-demon like system, the couplings has to be considered carefully.
\begin{figure}

  \includegraphics[width=2.40in,height=1.80in]{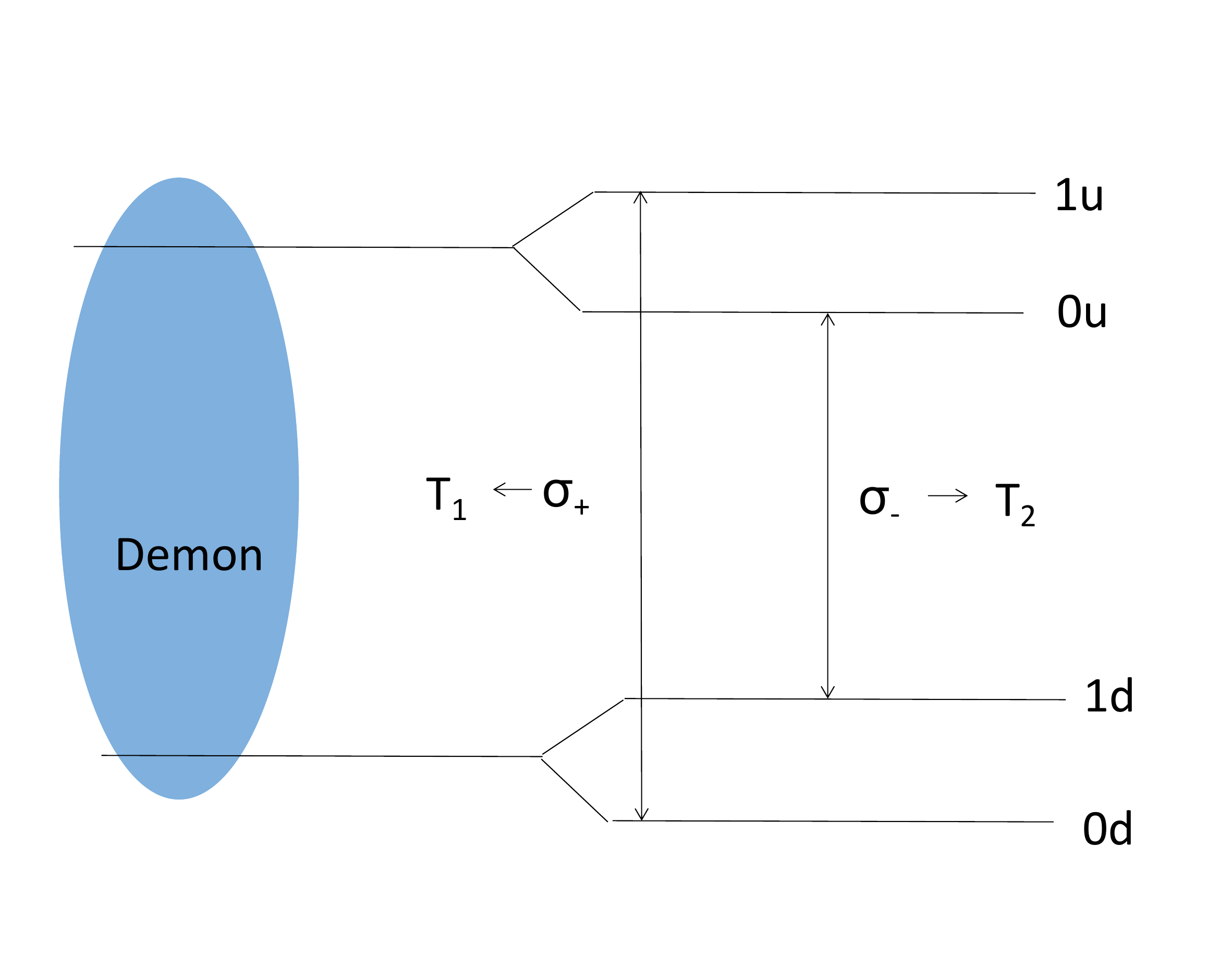}

  \caption{A rare earth ion system proposal for our Maxwell demon model.}
  \label{fig5}
\end{figure}

The Maxwell demon model we analyzed above does not require cumbersome calculations. In the model we discussed, we calculated the stable state of the demon, which depends on both information and the thermal reservoir. The directions of bits change and energy transportation are closely related. Under some conditions this may lead to erasing or refrigerating effects, while for other cases they simply reflects the energy flow in accordance with bits change.

We also checks the validity of the generalised the second of thermodynamics by calculating entropy change of the system, this can also be derived from the generalised Jarzynski equality. And the relative efficiency of erasing and refrigerating is maximum at reversible process only. In this model, Maxwell's demon has an essential role, it connects the information and the heat reservoir as a kind of information engine. Upon this model, we are going to study the situation when the demon and the bits are entangled. Because of the special properties of quantum information, the idea of using quantum mutual information as a source of heat engine is inspiring and it can lead to more understanding about quantum thermodynamics \cite{text4,text12,text14,text20,text21,text22}.

This work was supported by the 973 Program (Grant No. 2013CB921800), the NNSFC (Grant Nos. 11227901,
91021005, 11104262, 31470835, 21233007, 21303175, 21322305, 11374305 and 11274299), the ``Strategic Priority Research
Program (B)'' of the CAS (Grant Nos. XDB01030400 and 01020000)

\bibliographystyle{apsrev4-1}
\bibliography{New}

\end{document}